\pdfoutput=1
\documentclass[aps,prl,reprint,superscriptaddress,showpacs,preprintnumbers,amsmath,amssymb]{revtex4-1}
\usepackage[dvipdfmx]{graphicx}
\usepackage{dcolumn}   
\usepackage{bm}        
\usepackage{amssymb, amsmath, amsfonts}   
\usepackage{lipsum}        
\usepackage{color}
\DeclareGraphicsExtensions{{.pdf}}
\begin{document}
\newcommand{\noter}[1]{{\color{red}{#1}}}
\newcommand{\noteb}[1]{{\color{blue}{#1}}}
\newcommand{\field}{\left( \boldsymbol{r}\right)}
\newcommand{\paren}[1]{\left({#1}\right)}
\newcommand{\vect}[1]{\boldsymbol{#1}}
\newcommand{\uvect}[1]{\tilde{\boldsymbol{#1}}}
\newcommand{\vdot}[1]{\dot{\boldsymbol{#1}}}
\newcommand{\vder}{\boldsymbol{\nabla}}
%
\widetext
%
%
\title{
Glassy dynamics of a model of bacterial cytoplasm with metabolic activities}
\author{Norihiro Oyama}
\email{oyama.norihiro@aist.go.jp}
\affiliation{Mathematics for Advanced Materials-OIL, AIST, Sendai 980-8577, Japan}
\author{Takeshi Kawasaki}
\affiliation{Department of Physics, Nagoya University, Nagoya 464-8602, Japan}
\author{Hideyuki Mizuno}
\affiliation{Graduate School of Arts and Sciences, The University of Tokyo, Tokyo 153-8902, Japan}
\author{Atsushi Ikeda}
\affiliation{Graduate School of Arts and Sciences, The University of Tokyo, Tokyo 153-8902, Japan}
\affiliation{Research Center for Complex Systems Biology, Universal Biology Institute, University of Tokyo, Komaba, Tokyo 153-8902, Japan}

\date{\today}
\begin{abstract}
Recent experiments have revealed that cytoplasms become glassy when their metabolism is suppressed, while they maintain fluidity in a living state. The mechanism of this active fluidization is not clear, especially for bacterial cytoplasms, since they lack traditional motor proteins, which can cause directed motions. We introduce a model of bacterial cytoplasm focusing on the impact of conformational change in proteins due to metabolism. In the model, proteins are treated as particles under thermal agitation, and conformation changes are treated as changes in particle volume. Simulations revealed that a small change in volume fluidizes the glassy state, accompanied by a change in fragility, as observed experimentally. 
\end{abstract}
\maketitle
%
\emph{Introduction.}---
Active matter~\cite{Vicsek2012,Marchetti2013} encompasses a vast range of nonequilibrium systems such as living organisms~\cite{Angelini2011,Gerum2013,Lushi2014,Tennenbaum2016} and artificial objects including active Janus particles~\cite{Walther2013,Li2018} and anisotropic vibrated rods~\cite{Kudrolli2008} or discs~\cite{Deseigne2010}.
One distinguishing feature of active systems is that they are driven out of equilibrium without any external force or momentum injection.
Instead, they convert internal energy into mechanical work.
The understanding of dilute and moderately dense active matter has advanced significantly due to experiments, particle simulations, and continuum theory~\cite{Vicsek2012,Marchetti2013,Cates2014,Zottl2016,Oyama2016,Oyama2017,Oyama2017a}. 
However, the same level of understanding has not been achieved for dense active matter. 
Because many forms of biologically active matter are very dense, understanding them is necessary.
Such dense active matter exhibits glassy dynamics, which have similarities and interesting differences compared to those of passive systems~\cite{Zhou2009,Angelini2011,Parry2014,Garcia2015,Nishizawa2017,Nishizawa2017a}. 

To understand dense active matter, self-propelled particles (SPPs) have been studied. 
Numerical simulations established that the dynamics of SPPs become glassy when the density is sufficiently high~\cite{Henkes2011,Ni2013,Berthier2014,Levis2015,Szamel2015,Mandal2016,Flenner2016,Berthier2017,Mandal2017}. 
The signature of glassy dynamics strongly depends on the details of the system: interparticle interactions, temperature, density, and implementation of self-propulsion. 
The self-propulsion of particles can either suppress~\cite{Ni2013,Berthier2014,Mandal2016,Flenner2016,Berthier2017,Mandal2017} or promote~\cite{Flenner2016,Berthier2017} the slowing down of the dynamics. 
This self-propulsion also alters the fragility of the glassy dynamics, namely, the steepness of the increase in relaxation time against the control parameter: the system becomes either more fragile (super-Arrhenius)~\cite{Ni2013,Flenner2016} or stronger (Arrhenius)~\cite{Mandal2016,Mandal2017} with increasing departure from the equilibrium. 
In parallel to these simulation efforts, the mean-field theories of the glass transition~\cite{Berthier2011b} have been extended for dense SPPs. 
These theories successfully predict the emergence of glassy dynamics in dense SPPs and explain some of the findings of simulations~\cite{Berthier2013,Szamel2015,Liluashvili2017,Nandi2017,Nandi2018}. 
Moreover, they describe the violation of the fluctuation dissipation theorem and the emergence of the effective temperature in these systems~\cite{Berthier2013}. 

However, in reality, dense active matter is not restricted to SPPs~\cite{Bi2015,Tjhung2017a}.  
Here, we consider the interior of living cells, specifically the cytoplasm of bacteria (\emph{Escherichia coli}). 
Cytoplasm is a dense solution of macromolecules and organelles. 
Recent experimental measurements of diffusivity and viscosity revealed that bacterial cytoplasm behaves similarly to glass when adenosine triphosphate (ATP, the biological energy source) is depleted, while the dynamics are dramatically accelerated and the system behaves similarly to a fluid when ATP is supplied~\cite{Parry2014,Nishizawa2017,Nishizawa2017a}. 
Furthermore, this activeness was shown to alter the fragility of the system: the system becomes stronger when ATP is supplied~\cite{Nishizawa2017}. 
Quite interestingly, bacterial cytoplasm is different from dense SPPs in that the former does not have traditional motor proteins, which undergo directed motion and thus can be modeled as SPPs. 
Instead, as Parry et al. discussed, the metabolism of bacteria can cause non-SPP-type active perturbations, such as conformational changes in proteins, which may result in the fluidization of the glassy state~\cite{Parry2014}. 

At present, it is unclear if such conformational changes in proteins can indeed dramatically accelerate the dynamics of glassy systems. 
Recent studies showed that such conformational change can accelerate diffusion in dilute enzymatic solutions~\cite{Riedel2014,Mikhailov2015,Illien2017,Zhao2018}, but its impact on dense glassy systems is unknown. 
This work aims to answer this question using numerical simulations. 
We introduce a model of bacterial cytoplasm focusing on the effect of non-SPP-type activity, in which proteins are modeled as particles under thermal agitation and their conformational change is taken into account as a change in the volume of the particles. 
By means of a molecular dynamics (MD) simulation of the model, we show that small changes in particle volume drastically accelerate the dynamics and alter the fragility from fragile to strong, as observed in previous experiments~\cite{Nishizawa2017}. 

\emph{Numerical modeling.}---
Bacterial cytoplasm is too complex to handle numerically with a fully resolved description.
The size distribution of the constituents is very broad (ranging from the angstrom scale of water molecules and ions to the tens of nanometers scale of proteins), and proteins, the main constituents, have nonspherical shapes, deformability, and electrostatic and specific interactions.
In addition to these already-complicated passive features, the system shows active effects due to metabolic activities, e.g., conformational change in proteins.
Since recent experiments~\cite{Parry2014,Nishizawa2017} have shown that the introduction of activeness plays a crucial role in determining the glassy nature of the system,
we especially aim to clarify the effect of activeness on the glassy dynamics by discarding all other complexities.
As a simplified description of proteins,
we consider particles interacting via the harmonic repulsive potential: 
$V({r}_{ij}) = - \frac{\epsilon}{2\sigma_0^2}(\sigma_{ij}-r_{ij})^2 \Theta(\sigma_{ij}-r_{ij})$, where $r_{ij}$ is the distance between particles $i$ and $j$; $\sigma_{ij} = (\sigma_i + \sigma_j)/2$, with $\sigma_i$ being the diameter of particle $i$; and $\Theta(x)$ is the Heaviside step function. 
The dynamics are described by the overdamped Langevin equation, whose damping term models the contributions from the background solvent molecules:
\begin{eqnarray}
\eta \cfrac{\partial \boldsymbol{r}_i}{\partial t} = 
-\sum_{i\ne j}\frac{\partial V(r_{ij})}{\partial \boldsymbol{r}_i} +\boldsymbol{\xi}_i(t), \label{eom}
\end{eqnarray}
where $\boldsymbol{r}_i$ is the position of particle $i$, $\eta$ represents the damping coefficient and $\boldsymbol{\xi}_i(t)$ is the thermal agitation noise, which obeys the fluctuation dissipation theorem as $\langle \boldsymbol{\xi}_i(t)\boldsymbol{\xi}_j(t^{\prime})^T \rangle =2T\eta\delta_{ij}\boldsymbol{1}\delta(t-t^{\prime})$. 
We solve Eq.~(\ref{eom}) by commonly used MD simulation techniques with discretization of the time step $\Delta t$~\cite{Allen1989}. 

The effect of conformational change in constituent particles with ATP consumption is treated as an effective volume change in particles\cite{Togashi2019}.
We assume that each particle can take two states with different diameters: for the $i$th particle, the diameter is $(1+a)\sigma_i^0$ in one state and $(1-a)\sigma_i^0$ in the other, where $\sigma_i^0$ is the reference diameter of the $i$th particle and $a$ represents the amplitude of the active diameter change. 
Each particle is randomly transformed from one state to the other with the probability $r_a$ per unit time. 
This assumption is similar to the one employed to study dilute enzymatic solutions, where the stiffness of particles differs between two states~\cite{Illien2017}. 
To implement this process in the model, we introduce stochastic time dependence of the diameters of particles: for the $i$th particle, $\sigma_i(t) = \left[ 1+{a}\cdot(-1)^{n_i(t)} \right] \sigma_i^0$, where $n_i(t) = n_i^0 + \sum_{t=t_0}^t \Theta(\xi_i^t - r_{\rm a} \Delta t)$ is a time-dependent integer, $n_i^0$ is the initial value, and $\xi_i^t\in[0,1]$ is a uniform random number generated at each time step. 
Because this activeness works only as a change in the interaction parameter and neither external forces nor translational active motions are introduced, this model fully satisfies the requirement of non-SPP-type active matter.
The diameter change terminates in a single time step in the current model, which is natural because the time required for a single protein conformational change is very small compared to that for the diffusive dynamics of proteins ($\sim$ tens of ps)~\cite{Ishikawa2008}. 
For comparison, we also considered models with a finite duration of the diameter change and confirmed that the glassy dynamics of such a model are qualitatively the same as those of the present model (see Supplemental Material S1). 
Note that when we set $T=0$ and adopt sinusoidal time dependence of diameters, this model reduces to the model for the assembly of cells on cell sheets~\cite{Tjhung2017a}. 

Below, we focus on the two-dimensional version of this model with binary size dispersity of particles. 
The large and small particles' reference diameters are $\sigma_{\rm L}^0 = \sigma_0$ and $\sigma_{\rm S}^0 = 0.71\sigma_0$, and the proportion is $40\%:60\%$.
This system allows us to prepare a disordered configuration easily, and the initial states for simulations are all disordered unless otherwise stated.
The number of particles is $N=1000$. 
The packing fraction is $\varphi \approx 0.94$, although it slightly fluctuates over time (by $0.1\%$ at most) due to the fluctuation of particle diameters. 
The typical interval of the active size change $\tau_{\rm a}=2/r_{\rm a}$, during which particles experience one change cycle (two size changes) on average, is fixed at $\tau_{\rm a}=820\tau_0$, where $\tau_0=\eta\sigma_0^2/\epsilon$ is the time scale of the dissipation. 
{The following results do not qualitatively depend on changes in $\varphi$ and $\tau_{\rm a}$.}
Under these conditions, the control parameters of the present model are the temperature $T$ and the amplitude of active diameter change $a$ (we refer to this parameter as the \emph{activeness}). 
The system departs from equilibrium when $a >0$. 
We set the Boltzmann constant to unity, and we use $\sigma_0$, $\tau_0$ and $\epsilon$ as the units of the length, time and temperature, respectively.
{The Langevin equation is integrated numerically by using Euler-type discretization~\cite{Allen1989}
with the time increment $\Delta t=0.1$.
The results are robust against a change in $\Delta t$ by a factor of ten.}

\begin{figure}
\includegraphics[width=\linewidth]{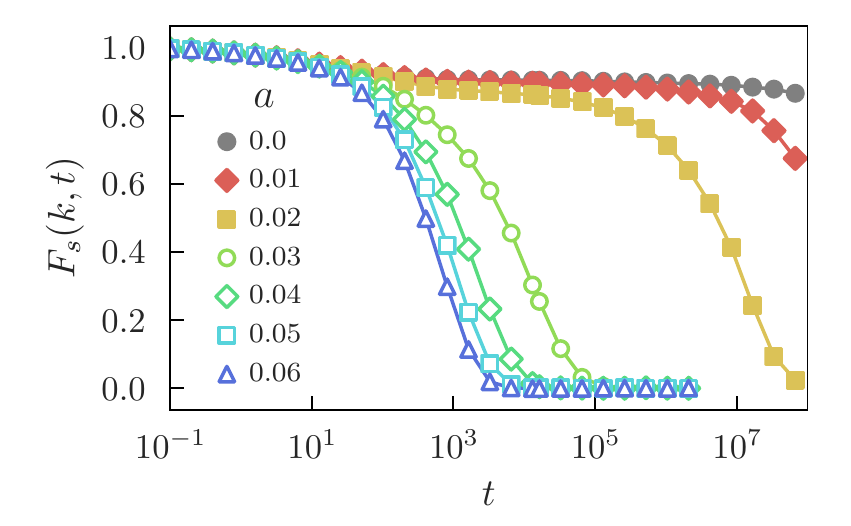}
\caption{(Color online)
  Self-part of the intermediate scattering function $F_{\rm s}(k, t)$ with $k\simeq 2\pi$ at $T = 0.003$.  
  Different symbols (colors) indicate different values of the diameter change $a$.
\label{fig:fig1}}
\end{figure}
\emph{Active fluidization due to diameter change.}---
We first measure the self-part of the intermediate scattering function (SISF) as $F_{\rm s}(k,t) = \langle \hat{F}_{\rm s}(k,t) \rangle$, where $\hat{F}_{\rm s}(k,t) = \frac{1}{N} \sum_{j=1}^N \text{exp}[i\boldsymbol{k}\cdot(\boldsymbol{r}_j(t) - \boldsymbol{r}_j(0))]$, 
$\boldsymbol{k}$ and $k\equiv |\boldsymbol{k}|$ stand for the wave vector and its magnitude, respectively, and angular brackets indicate the time average.
We set $k \simeq 2\pi$ to study the relaxation dynamics at the particle scale.
Fig. \ref{fig:fig1} shows the results for $T=0.003$. 
At this temperature, the SISF of the passive system ($a=0.0$) decreases slightly over a short time, the so-called $\beta$ relaxation, and then remains at a plateau for a long time, which is a basic signature of the dynamics of glass states. 
In contrast, the dynamics are drastically accelerated with an increasing amplitude of active diameter change $a$. 
The SISF exhibits slight decay, even at $a = 0.01$, and decays to zero within the observation time window when $a \gtrsim 0.02$.
Note that $a = 0.02$ corresponds to a $4\%$ diameter change because the ratio of diameters in the two states is $(1+a)/(1-a)$. 

We now argue that the fluidization due to such small diameter changes is a characteristic of the glassy state. 
We performed similar simulations but starting from the triangle lattice structure of particles (see Supplemental Material S2).  
We find that the fluidization of this crystalline packing takes place at only $a \gtrsim 0.1$, namely, a $20\%$ diameter change. 
This threshold value is clearly much larger than that for the fluidization in the glassy state. 
Therefore, we conclude that the glassy state is much weaker against particle-scale perturbations than the crystalline state, and a small diameter change is sufficient to drastically accelerate the dynamics of and fluidize the glassy state.

 
%
\begin{figure}
\includegraphics[width=\linewidth]{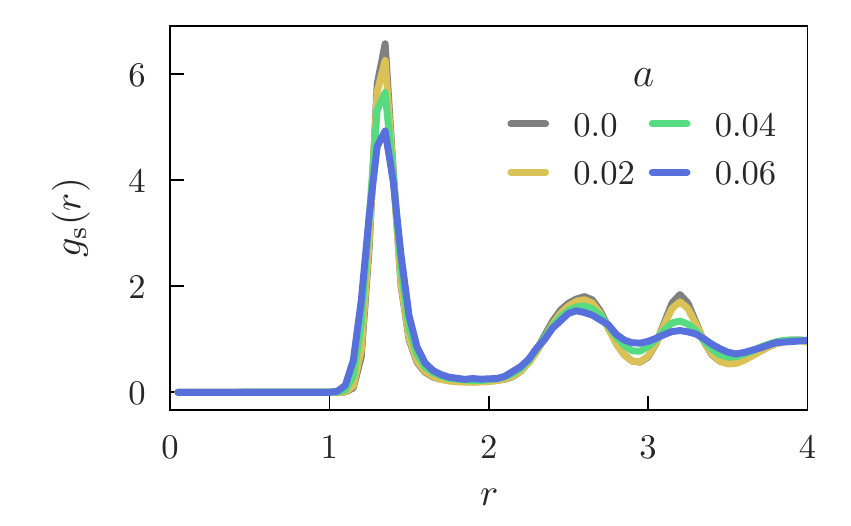}
\caption{\label{fig:fig2}
(Color online) 
Pair distribution function of small particles $g_{\rm s}(r)$ as a function of interparticle distance $r$.
As is the case for Fig.~\ref{fig:fig1}, all results are from the simulation with $T=0.003$.
Different colors represent different values of $a$, as listed in the legend.
}
\end{figure}
In the case of SPPs, the first peak of the radial distribution function is enhanced by the introduction of activeness. 
This clustering tendency has been reported rather universally in dense SPPs~\cite{Flenner2016,Mandal2016,Berthier2017}, while other features strongly depend on the details of the systems. 
In Fig.~\ref{fig:fig2}, we present the radial distribution function of small particles $g_{\rm s}(r)$ for $T=0.003$ and several values of $a$.
As the activeness $a$ is increased, the heights of peaks become lower monotonically; thus, the clustering tendency due to activeness is absent in this model.  
We also calculated the spatial correlation function of the longitudinal velocity: this correlation is known to strengthen with an increase in the activeness of dense SPPs~\cite{Flenner2016, Berthier2017}.  
We found that this spatial correlation function is structureless in our model, as in equilibrium fluids (see Supplemental Material S3).
Therefore, the structuring tendency due to activeness is absent in the present model, which is in contrast to SPPs. 

%
\begin{figure}
\includegraphics[width=\linewidth]{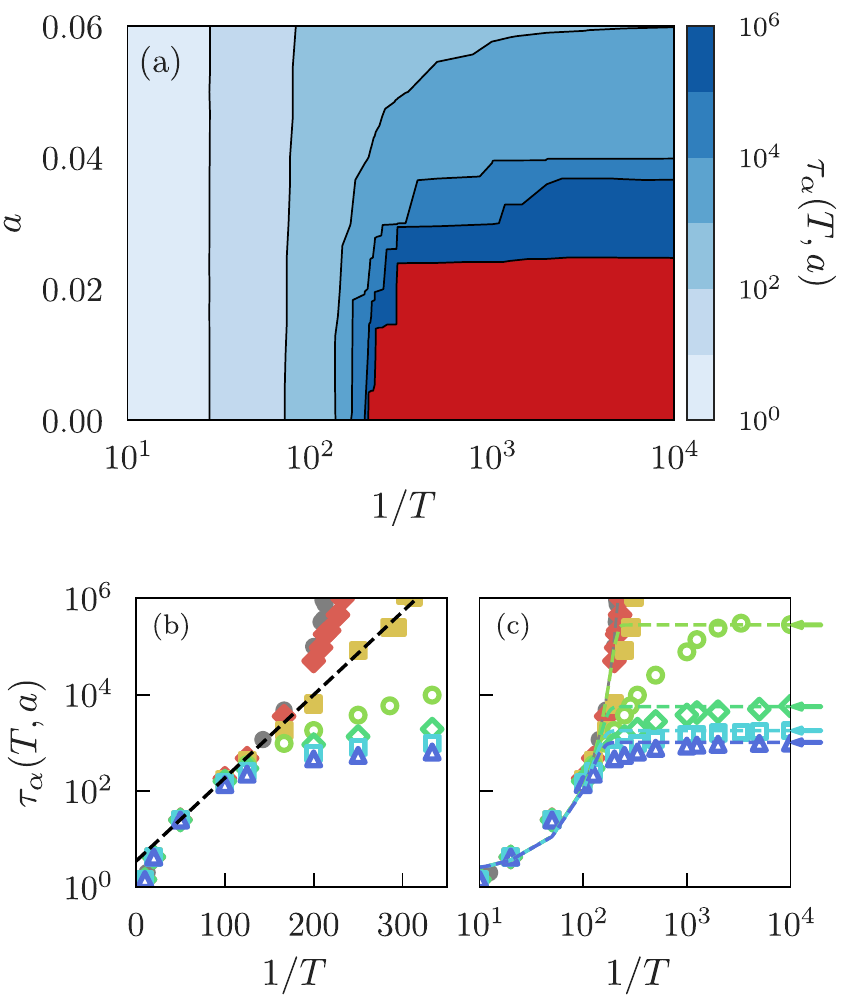}
\caption{\label{fig:fig3}
(Color online) 
(a) Contour map of the relaxation time $\tau_\alpha$ against the inverse temperature $1/T$ and the diameter change $a$. 
The region with $\tau_{\alpha} \geq 8\times 10^5$ is filled with red. 
(b) Semilogarithmic plot of the relaxation time $\tau_\alpha$ as a function of the inverse temperature $1/T$ for various values of $a$.
The black dashed line indicates the Arrhenius behavior.  
(c) Same as the bottom left, but showing a double-logarithmic plot.
The arrows to the right indicate the relaxation time at $T=0$, and the dashed lines indicate $\tau_{\alpha}^{\rm ind}(T,a)$, defined in Eq.~(\ref{model}). 
In (b) and (c), the colors and symbols indicate $a$, as listed in Fig.~\ref{fig:fig1}.
}
\end{figure}
\emph{Relaxation time for the glassy dynamics.}---
We then repeated the calculations of the SISF for various state points.  
Fig.~\ref{fig:fig3}~(a) shows the relaxation time $\tau_{\alpha}$ defined by $F_{\rm s}(k \simeq 2\pi, \tau_{\alpha})=1/e$ against temperature $T$ and the activeness $a$. 
The system maintains fluidity for any temperature when $a \gtrsim 0.03$. 
Glass states emerge only when the temperature is low enough, i.e., $T \lesssim 10^{-2}$, and the diameter change is small enough, i.e., $a \lesssim 0.02$. 
For more quantitative analysis, we also plot $\tau_{\alpha}$ against $1/T$ for various values of $a$ in panels (b) and (c) of Fig.~\ref{fig:fig3}. 
The semilogarithmic plot in Fig.~\ref{fig:fig3}~(b) shows that $\tau_{\alpha}$ increases with $1/T$ in a super-Arrhenius manner in the passive system ($a = 0$), which is a hallmark of fragile glass formers\footnote{Note that the fragility of the passive case, $a=0.0$, is expected to depend on the choice of $\varphi$~\cite{Berthier2009}.
}.
However, the increase in $\tau_{\alpha}$ becomes weaker with increasing $a$. 
$\tau_{\alpha}$ for $a = 0.01$ starts to deviate from the passive ones only at a very low temperature, $T \lesssim 0.005$.
The deviation appears at a higher temperature for $a = 0.02$, and $\tau_{\alpha}$ follows the Arrhenius law in a wide temperature region, which is a hallmark of strong glass formers. 
For larger values of $a$, the deviation appears at a much higher temperature, $\tau_{\alpha}$ follows the Arrhenius law at intermediate temperatures, and $\tau_{\alpha}$ finally seems to saturate at very low temperatures. 
This saturation becomes clear in the double-logarithmic plot (Fig.~\ref{fig:fig3}~(c)). 
At $T \to 0$, $\tau_{\alpha}$ converges to well-defined values.
These limiting values coincide with the $\tau_{\alpha}$ of purely athermal systems, which are obtained from independent simulations at $T=0$ and indicated by arrows. 
Therefore, the relaxation time $\tau_{\alpha}(T,a)$ in the present model smoothly connects the purely thermal results $(T \neq 0, a = 0)$ at a high temperature and purely athermal results $(T = 0, a \neq 0)$ at a lower temperature, and Arrhenius behaviors emerge at the intermediate temperatures.

To gain more insight at the intermediate temperatures, we test the following simple assumption: if the thermal and athermal relaxations take place independently, the relaxation time (denoted by $\tau_{\alpha}^{\rm ind}(T,a)$) can be expressed as 
\begin{eqnarray}
\frac{1}{\tau_{\alpha}^{\rm ind}(T,a)} = \frac{1}{\tau_{\alpha}(T,0)} + \frac{1}{\tau_{\alpha}(0,a)}, \label{model} 
\end{eqnarray}
where $\tau_{\alpha}(T,0)$ and $\tau_{\alpha}(0,a)$ are the relaxation time of purely thermal and athermal systems, respectively. 
We evaluated $\tau_{\alpha}^{\rm ind}(T,a)$ using $\tau_{\alpha}(T,0)$ and $\tau_{\alpha}(0,a)$ obtained from the independent simulations at $a=0$ and $T=0$, respectively, and plotted them as dashed lines in Fig.~\ref{fig:fig3}c. 
Clearly, $\tau_\alpha(T,a)$ is comparable to $\tau_{\alpha}^{\rm ind}(T,a)$ at high and low temperatures but significantly smaller than  $\tau_{\alpha}^{\rm ind}(T,a)$ at the intermediate temperatures. 
This result suggests that the thermal and athermal relaxations are not independent but that the interplay between them significantly accelerates the dynamics in this region. 
Such interplay has been observed in sheared glasses, where the yielding process is assisted by thermal activation such that the shear stress is noticeably decreased at finite temperature~\cite{Chattoraj2010}. 
We speculate that a similar mechanism accelerates the dynamics in the present model at the intermediate temperatures. 

\begin{figure}
\includegraphics[width=\linewidth]{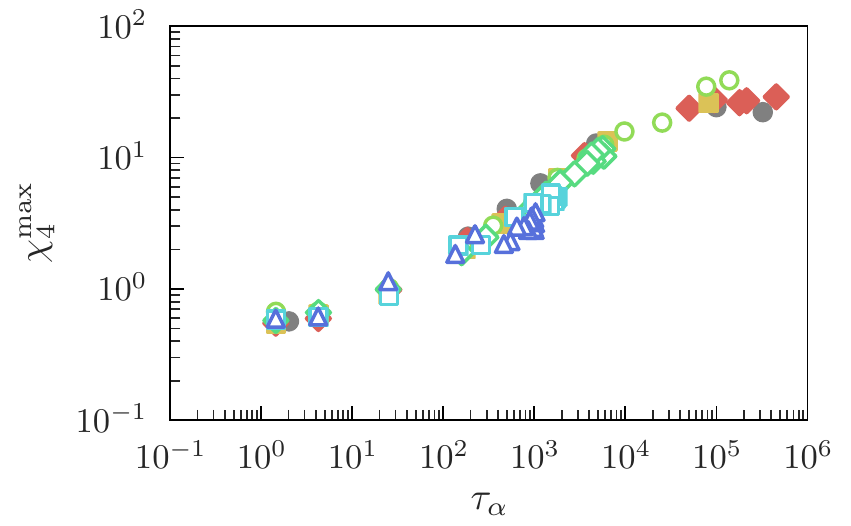}
\caption{\label{fig:fig4}
(Color online) 
Maximum value of $\chi_4(k,t)$ against the relaxation time $\tau_{\alpha}$ for various values of temperature $T$ and diameter change $a$. 
Colors and symbols indicate $a$, as listed in Fig.~\ref{fig:fig1}.
Only the data obtained from sufficiently long simulations ($50\tau_\alpha\le \tau_{\rm sim}$) are shown, where $\tau_{\rm sim}$ is the total simulation time.
}
\end{figure}
\emph{Dynamic heterogeneity.}---
We also quantified the heterogeneity in the dynamics by measuring nonlinear susceptibility: $\chi_4(k,t) = N [\langle \hat{F}_{\rm s}(k,t)^2 \rangle - \langle \hat{F}_{\rm s}(k,t) \rangle^2]$. 
We set $k \simeq 2\pi$, as in the case of the SISF. 
In all cases of $T$ and $a$, $\chi_4(k,t)$ first increases with time, reaches a maximum at $t \sim \tau_{\alpha}$, and then decreases. 
In Fig.~\ref{fig:fig4}, we plotted the maximum value $\chi_4^{\rm max}$ as a function of $\tau_{\alpha}$. 
The dynamics become more heterogeneous as the relaxation dynamics slow, as in the case of passive glass-forming systems. 
Interestingly, $\chi_4^{\rm max}$ at various $T$ and $a$ values collapses into a single master curve of relaxation time. 
This behavior is again similar to that of sheared glasses, where the dynamic correlation lengths at various temperatures and shear rates are collapsed into a single master curve of  relaxation time~\cite{Yamamoto1998}. 
This consistency supports the similarity of the relaxation mechanisms between our model and sheared glasses. 
Note that a similar collapse was also observed in dense SPPs when the effects of the short-term dynamics were renormalized~\cite{Flenner2016}. 

\emph{Summary.}---
In this work, we composed a model of bacterial cytoplasm focusing on the effect of the conformational change in proteins, which is taken into account as a change in the volume of particles. 
By means of MD simulation, we numerically proved that a very small diameter change, $6\%$, is sufficient to drastically accelerate the dynamics and fluidize the glassy system. 
This fluidization was accompanied by a change in fragility: the actively fluidized system was more Arrhenius-like. 
All these results are consistent with the experimental observations of active fluidization of bacterial cytoplasm. 
We also showed that the relaxation dynamics under active diameter changes have several similarities to those of sheared glasses. 
More information about this similarity can be obtained by analyzing the change in the potential energy landscape over the course of diameter change, as was analyzed for the yielding of glasses under shear~\cite{Maloney2006}.  
We are now working in this direction. 
Another interesting research direction is to extend the present model to take into account more realistic interactions between proteins, such as attractive and specific interactions, to directly compare the simulation results with the results of experiments. 

{We thank D. Mizuno, K. Nishizawa and H. Itoh for useful discussions.
  This work was financially supported by KAKENHI grants
  (nos. 16H04025, 16H04034, 16H06018, 17H04853, 17K14369, 18H01188, 
  and 19K14670) and partially supported by the Asahi Glass Foundation.}

	
%
%

%
%
\end{document}